\documentclass[slac_one]{revtex4}
\usepackage{graphicx}
\usepackage{fancyhdr}
\newcommand{\F}{$ F_{2}(x,Q^2)\:$} 

\newcommand{\Fc}{$ F_{2}\,$}

\newcommand{\FLc}{$ F_{L}\,$}

\newcommand{\pdff}{$\partial F_{2} / \partial \ln Q^{2}\,$ }
\begin{document}

\title{\boldmath First Measurements of $F_L$ at Low Bjorken $x$} 

%

\author{M. Klein (for the H1 and ZEUS Collaborations)}
\affiliation{University of Liverpool, Physics Department, L69 7ZE, UK}

\begin{abstract}
The first measurements  of the longitudinal proton structure
function \FLc in deep inelastic positron-proton scattering at low Bjorken $x$
are described.
Theoretical predictions in higher order QCD and using dipol model
calculations are consistent with the measurements.
The data were taken by the HERA experiments H1 and ZEUS  in a series
of runs with different proton beam energies in 2007.
\end{abstract}

\maketitle

\thispagestyle{fancy}

\section{\boldmath  $F_L$ at HERA }
The measurement of the longitudinal proton structure function
 had long been recognised as an important part
of the HERA programme. 
The first estimates on the measurement accuracy 
of \FLc were presented more than 20 years
ago within the 1987 workshop on HERA physics~\cite{hera87}.
In 1996 both experiments had pursued quite some preparations
for the measurement of \FLc  with lowered proton beam
energies~\cite{lbagmk}. The measurement was yet postponed
by  $10$ years, basically for clarification of anomalies
in the data at  large scales, which were possibly related to leptoquarks and to
peculiar events with isolated leptons accompanied by large
missing transverse momentum~\cite{rikmax}. In 2004, at the annual DIS workshop
the interest in a measurement of \FLc was renewed~\cite{dis04}.
 Meanwhile the HERA collider
had successfully completed a luminosity upgrade program which
for the \FLc measurement was particularly important
as the luminosity decreases $\propto E^2_p$ with the proton beam
energy. In 2005 (2006) the H1 (ZEUS) experiments did request 
to  lower the proton beam energy in order to measure \FLc
prior to the termination of HERA which had been announced
for June 2007.  Both experiments had performed essential upgrades
on their detector, H1 with the SpaCal calorimeter, a Backward Silicon
Tracker and improved trigger and backward chambers, ZEUS with
a Micro Vertex Silicon Detector and a renewed electron tagging device.
Both experiments therefore were well prepared for a measurement
of \FLc. The understanding of the machine had reached a 
most impressive level~\cite{ferdi}, as became clear during the low energy run
when the optics, the polarisation (for HERMES), the luminosity and
the overall operation efficiency were all up to or even beyond
expectation. The data have promptly been analysed by both collaborations.

Data were taken with proton beam energies of
$E_p = 460$\,GeV and $575$\,GeV and compared with existing
high energy data at $920$\,GeV while the positron beam
energy was kept constant, $E_e=27.5$\,GeV.
At the DIS08 workshop first observations were reported
by both Collaborations~\cite{dis08}:
the H1 Collaboration had released its first data for
both lower proton energies, using the backward calorimeter
SpaCal and the central Liquid Argon Calorimeter (LAr).
The ZEUS Collaboration did present a first analysis on their
$460$\,GeV data set.  The present report is based
on updates by H1 and ZEUS prepared for the ICHEP conference:
the SpaCal data of H1 were published~\cite{h1flpl} while
ZEUS did present a combined \FLc 
analysis based on also the $575$\,GeV data.

The measurements presented to ICHEP08 were the first ever 
direct determinations of \FLc 
in the region of low Bjorken $x$. Data at large $x$ had been
obtained by a number of fixed target lepton-proton scattering 
experiments~\cite{rdata}, beginning with the discovery at SLAC that 
the ratio $R \simeq F_L/(F_2 -F_L)$ was close to zero which
was a most convincing evidence for quarks to be spin $1/2$ 
fermions. While subsequent measurements with muon 
and neutrino beams
mostly observed small values of $R$,  indirect measurements
of \FLc by H1~\cite{fl94} did hint to  larger values of the
longitudinal proton structure function at low $x$. This had been
expected, in the region of deep inelastic scattering, from 
the large size of the gluon distribution in this range. While
$R$ at large $x$ is a measure of the quark's spin, at low $x$
it rather quantifies the dynamics of gluon interactions
and as such is of particular theoretical interest~\cite{rtjb}.
%
\section{\boldmath Measurement of $F_L$}
The inclusive, deep inelastic electron-proton scattering cross section at low $Q^2$,
written in its reduced form,
\begin{equation}
 \sigma_r = \frac{d^2\sigma}{dxdQ^2} \cdot    \frac{Q^4 x} {2\pi \alpha^2 Y_+} =  
    F_2(x,Q^2) - \frac{y^2}{Y_+} \cdot F_L(x,Q^2)
       \label{sig}
  \end{equation}  
is defined by two proton structure functions, \Fc and \FLc, where
$Q^2$ is the negative four-momentum transfer squared, $y$ the inelasticity
$y=Q^2/sx$ with $Y_+ = 1+ (1-y)^2$,
 $s=4 E_e E_p$ the centre of mass energy squared given
by the positron and the proton beam energies and $\alpha$ is
the fine structure constant.  Apparently, $\sigma_r$ is a direct measure
of \Fc apart from a limited region of high $y$, corresponding to small
$x$, where the contribution of \FLc may be sizeable. Disentangling the
two cross section terms requires to measure $\sigma_r$ at fixed $x$
and $Q^2$ at various beam energies. The analysis of the reduced
cross section determines \Fc as the intercept, at $y=0$, and \FLc
as the negative slope of its linear dependence on $y^2/Y_+$.  The variation
in $y$ was achieved at HERA by comparing high  statistics data 
at highest energy with about $13$\,pb$^{-1}$ of  data at $460$\,GeV
and $7$\,pb$^{-1}$ at $575$\,GeV. The values of $E_p$ had been
chosen for about equidistant separation of the measurements in
$y^2/Y_+$. The low energy runs took place from March to 
June 2007, with only a few days of setup time for the machine.

 The inelasticity is determined by the
energy and polar angle of the scattered positron,
$y = 1 -(E_e'/E_e) \sin^2(\theta_e/2)$, 
which at large scattering angles, corresponding to
small $Q^2 = 4 E_e E_e' \cos^2(\theta_e/2)$,
reduces to  $y \simeq 1- E_e'/E_e$.  If therefore one intends to
measure the cross section at high $y \simeq 0.9$ one needs to
master the trigger rate and the electron identification
at small energies of a few GeV.   Such energies are
deposited much more frequently by hadrons from 
photoproduction processes, for which $Q^2 \simeq 0$,
than by the genuine scattered DIS positron. Removal
of the photoproduction background constitutes the
major challenge of the measurement of \FLc at HERA.

There are two ways to control the photoproduction ($\gamma p$) 
background: part of the $\gamma p$ events is uniquely
identified by  tagging the scattered positron near to the
beam axis in  tagging calorimeters downstream the
$e^+$ beam. That measurement can be used to tune the
Monte Carlo simulation of the background, subject to uncertainties
of the tagger acceptance and its extrapolation. In a further 
method one can employ the charge symmetry of the background.
While a DIS positron carries the lepton beam charge, energy depositions
due to hadrons from $\gamma p$ processes are charge symmetric,
apart from a small proton-antiproton cross section difference
one can correct for.  With a tracking detector in front of the
calorimeter one may determine the charge of the DIS
lepton candidate and subtract the wrong charge 
signal statistically. ZEUS has subtracted the
background with a tagger based simulation, for $E_e' > 6$\,GeV,
in which subprocesses simulated in PYTHIA are weighted
using ZEUS $\gamma p$ cross section data.
H1 has primarily used the charge measurement and included data
for $E_e' > 3$\,GeV.   
Both analyses require a track detector signal associated
to the calorimeter energy deposition which reduces
the contribution from neutral particles, in particular
 of the electromagnetic energy deposition 
from $\pi_0 \rightarrow \gamma \gamma$.
Energy momentum
conservation is imposed requiring $E -p_z > 35~(42)$\,GeV, for H1 (ZEUS),
which further reduces the $\gamma p$ background and higher order
QED effects. 

Extraction of $F_L(x,Q^2)$ for each $x,Q^2$ is achieved by fitting the cross section
measurements as a function of $y^2/Y_+$. 
Utilising the large range in inelasticity $y$, both the ZEUS and the H1
analyses at this stage have renormalised the measured cross sections
at low $y$ where at each energy, for fixed $x$ and $Q^2$, the
same \F measurement is provided by the measurements of 
$\sigma_r$ from the up to three data sets. The systematic
errors  are taken into account source by source in an offset method
which leads to a reduction of their effect on the final measurement.
\section{Results}
\begin{figure*}[h]
\centering
\includegraphics[width=95mm]{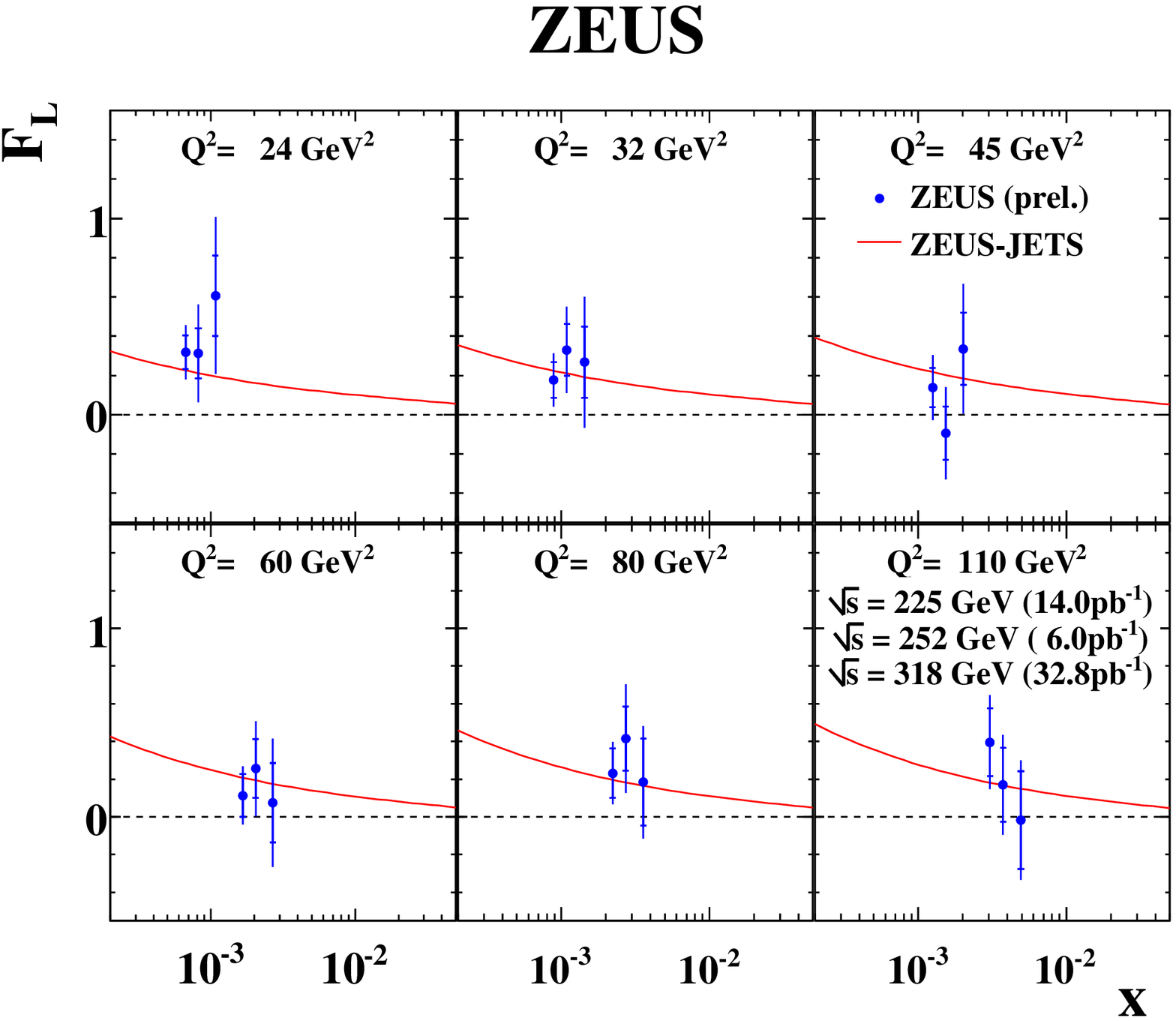}
\caption{Measurement of $F_L(x,Q^2)$ by ZEUS compared
with an NLO QCD prediction based on other ZEUS data.}
\label{figflzeusprel}
\end{figure*}

The resulting measurement of ZEUS, for $24 \leq Q^2 \leq 110$\,GeV$^2$,
is illustrated  in  Figure~\ref{figflzeusprel}.
The preliminary result is compared with a ZEUS QCD fit to NLO
to previous ZEUS DIS and jet data. At low $x$ and not too small
$Q^2$ the prediction on \FLc relies nearly completely on
the behaviour of the gluon distribution. That is determined
at low $x$ by the scaling violations of \F. The consistency
of the theoretical calculation on \FLc \cite{qcdfl} 
with the data is therefore
a non-trivial test of QCD to high orders.

\begin{figure*}[h]
\centering
\includegraphics[width=95mm]{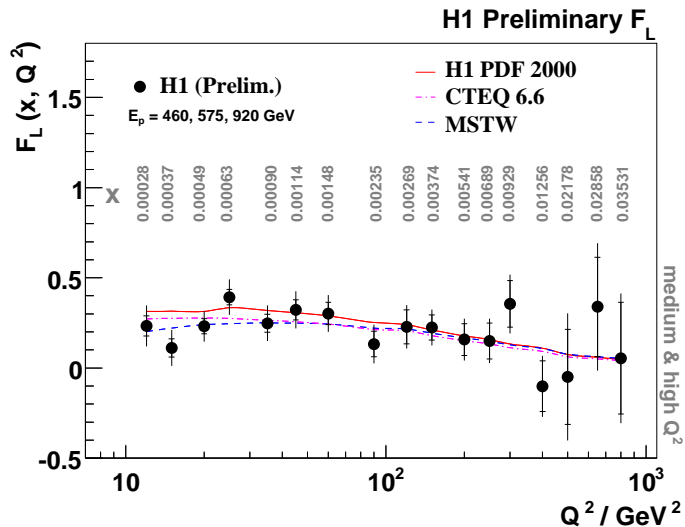}
\caption{Measurement of $F_L(x,Q^2)$ by H1
compared with different predictions in higher order QCD.} \label{figflh1prel}
\end{figure*}

For H1, a combination of the published, medium $Q^2$,
result based on SpaCal data~\cite{h1flpl},
for $12 \leq Q^2 \leq 90$\,GeV$^2$,
with a preliminary result based on LAr data at higher $Q^2$,
$35 \leq Q^2 \leq 800$\,GeV$^2$,
 is shown in Figure~\ref{figflh1prel}. The overlap of the cross section
 measurements with the two calorimeters at different $E_p$ 
reduces the uncertainty on the \FLc measurement, compared to~\cite{h1flpl}.
At each value of $Q^2$ one value on \FLc is derived which results from
an average of typically three data points adjacent in $x$. The data are
compared with three QCD predictions at higher orders, from the H1PDF2000
fit and recent CTEQ and MSTW fits on the parton distributions. 
The QCD predictions describe the measurements well. There are two additional
observations worth mentioning here: i) at low $x$ one expects
that \pdff /2 \FLc is approximately one, because both the derivative \cite{prytz}
and \FLc \cite{hera87} to LO are proportional to the dominating gluon density.
From the published 
H1 data on the derivative and on \FLc, for $Q^2$  between $10$ and $30$\,GeV$^2$,
this ratio is determined to be  \pdff /2 \FLc$ = 1.09 \pm 0.13 (stat) \pm 0.20 (syst)$
and indeed consistent with one;
ii) the measurement of \FLc is also consistent with predictions
within the dipole model, e.g. \cite{schild}.  In particular, no violation is observed of the
constraint $F_L \leq 0.27 F_2$ which reflects the 
wave function relations in the dipole model~\cite{nacht}.

The results presented here are the first, not the final measurements of
$F_L$ at low $x$ at HERA. There are a number of improvements being 
investigated regarding the extension of the kinematic
range to lower $Q^2$, to lower and possibly higher $y$, the understanding of the 
systematic errors, of the $\gamma p$ background in particular, or the
unfolding of \Fc and \FLc from the measured cross sections. One
can thus expect the accuracy of \FLc to be enhanced in the future.

\begin{acknowledgments}
The measurement of \FLc became possible thanks to the engagement
and competence of the HERA machine crew and the continued
support of the collider experiments ZEUS and H1 by the DESY
directorate. 
\end{acknowledgments}


\begin{thebibliography}{9}   

%
\bibitem{hera87}
A.M.\,Cooper-Sarkar et al.,
Z.\,Phys. {\bf C39} (1988) 28, also
 HERA Workshop 1987, Proceedings  Vol 1, p.231; \\
J.\,Blumlein et al., 
PHE-88-01, Vol 1, p.67, Hamburg 1987, ed. R. Peccei.
%
\bibitem{lbagmk}
  L.A.T.\,Bauerdick, A.\,Glazov and M.\,Klein,
  HERA Physics Workshop, 1996/97, Proceedings, eds. A.\,De\,Roeck and R.\,Klanner
  [hep-ex/9609017].
%
\bibitem{rikmax}
For a recent overview on HERA Collider physics see: \\
M.\,Klein and R.\,Yoshida, Prog.\,Part.\,Nucl.\,Phys. {\bf 61} (2008) 343.
%
\bibitem{dis04}
%
A.\,Martin,
 Proceedings DIS04 Workshop, Vol.
1, p.146; \\
M.\,Klein, 
Vol.1, p.309,  
eds. D. Bruncko,  J. Ferencei and P. Strizenec [http://www.saske.sk/UEF/OSF/DIS/]
%
\bibitem{ferdi}
F.\,Willeke, ``Prospects for Operating HERA with Lower Proton Energy",
informal memo, 15th of September 2005, unpublished.
%
\bibitem{dis08}
B.\,Antunovi\'c and V.\,Chekelyan (H1) and D.\,Kollar (ZEUS),
Proceedings DIS2008 Workshop, London, UK, April 2008, 
eds.  R.\,Devenish, J.\,Ferrando and M.\,Wing, to appear online.
%
\bibitem{h1flpl}
F.D.\,Aaron et al., H1 Collaboration, Phys. Lett. {\bf B665} (2008) 139.
%
\bibitem{rdata}
 J.J.\,Aubert et al., EMC Collaboration, 
 Phys. Lett. {\bf B121} (1983) 87; \\
A.C.\,Benvenuti et al., BCDMS Collaboration, 
  Phys. Lett. {\bf B223} (1989) 485; \\
L.W.\,Whitlow  et al., Phys. Lett. {\bf B250} (1990) 193; \\
M.\,Arneodo et al., NMC Collaboration,
 Nucl. Phys. {\bf B483} (1997) 3.
%
\bibitem{fl94}
 C.\,Adloff et al., H1 Collaboration,  Phys. Lett. {\bf B393} (1997) 452; \\
  C.\,Adloff et al., H1 Collaboration,  Eur. Phys. J. {\bf C21} (2001) 33.
%
\bibitem{rtjb}
For example see:
J.\,Blumlein et al.,    Nucl.\,Phys. {\bf B755} (2006) 272;   \\
A.D.\,Martin, W.J.\,Stirling and R.S.\,Thorne,
  Phys.\,Lett.  {\bf B635} (2006) 305.
%
\bibitem{qcdfl}
E.B.\,Zijlstra and W.\,van\,Neerven, Nucl.\,Phys. {\bf B383} (1992) 525; \\
S.A.\,Larin and J.A.M.\,Vermaseren, Z.\,Phys. {\bf C57} (1993) 93.
%
\bibitem{prytz}
K.\,Prytz, Phys.\,Lett. {\bf B311} (1993) 286.
%
\bibitem{schild}
M.\,Kuroda and  D.\,Schildknecht, 
 arXiv:0806.0202 [hep-ph].
%
\bibitem{nacht}
C.\,Ewerz and O.\,Nachtmann, Ann.\,Phys. {\bf 322} (2007) 1635.
%
\end{thebibliography}
\end{document}